\documentclass[apj]{emulateapj}
\usepackage{apjfonts}
\usepackage{graphicx}
\usepackage{psfig}
\slugcomment{{\sc Submitted to ApJ:} August 13, 2009 }
\newcommand{\begit}{\begin{itemize}}
\newcommand{\enit}{\end{itemize}}
\newcommand{\begen}{\begin{enumerate}}
\newcommand{\enen}{\end{enumerate}}
\newcommand{\p}{\partial}    
 
\newcommand{\beq}{\begin{equation}}
\newcommand{\eeq}{\end{equation}}
\newcommand{\beqa}{\begin{eqnarray}} 
\newcommand{\eeqa}{\end{eqnarray}} 
\newcommand{\bnabla}{\mbox{$\nabla$}}

\begin{document}

\title{Slow Diffusive Gravitational Instability Before Decoupling}

\author{Todd A.~Thompson\altaffilmark{1}}

\affil{Department of Astronomy and Center for Cosmology \& Astro-Particle Physics \\ 
The Ohio State University, Columbus, Ohio 43210 \\
thompson@astronomy.ohio-state.edu}

\altaffiltext{1}{Alfred P.~Sloan Fellow}

\begin{abstract}

Radiative diffusion damps acoustic modes at large comoving 
wavenumber ($k$) before decoupling (``Silk damping'').  
In a simple WKB analysis, neglecting moments of the temperature distribution
beyond the quadrupole (the tight-coupling limit), damping appears in the acoustic mode 
as a term of order $ik^2\dot{\tau}^{-1}$, where $\dot{\tau}$ is the scattering 
rate per unit conformal time.  Although the Jeans instability
is stabilized on scales smaller than the adiabatic Jeans length,
I show that the medium is linearly unstable to first order in $\dot{\tau}^{-1}$ to
a slow diffusive mode.  At large comoving wavenumber, 
the characteristic growth rate becomes independent of spatial scale and
constant: $(t_{\rm KH}\,a)^{-1}\approx(128\pi G/9\kappa_Tc)(\rho_m/\rho_b)$, 
where $a$ is the scale factor, $\rho_m$ and $\rho_b$ are the matter and baryon 
energy density, respectively, and $\kappa_T$ is the Thomson opacity.  
This is the characteristic timescale for a fluid parcel to 
radiate away its total thermal energy content at the Eddington limit,
analogous to the Kelvin-Helmholz (KH) timescale for a radiation pressure-dominated
massive star or the Salpeter timescale for black hole growth.
Although this mode grows at all times prior to decoupling and on
scales smaller than roughly the horizon, 
the growth time is long, about 100 times the age of 
the universe at decoupling.  Thus, it modifies
the density and temperature perturbations on small scales only
at the percent level.  The physics of this mode in the 
tight-coupling limit is already accounted for in the popular codes
{\tt CMBFAST} and {\tt CAMB}, but is typically neglected in analytic
studies of the growth of primordial perturbations.  The goal of this 
work is to clarify the physics of this diffusive instability
in the epoch before decoupling, and to emphasize that the
universe is formally unstable on scales below the horizon,
even in the limit of very large $\dot{\tau}$.
Analogous instabilities that might 
operate at yet earlier epochs are also mentioned. \\

\end{abstract}

\keywords{cosmic microwave background --- 
cosmology : theory --- large-scale structure of universe}

\section{Introduction}
\label{section:introduction}

In standard analytic treatments of perturbation theory in 
the epoch near radiation-matter equality the tight coupling
approximation is employed, which amounts to neglecting moments 
of the temperature distribution beyond the quadrupole.  In this 
limit, the scattering rate per unit conformal time $\dot{\tau}$ 
is very large and, neglecting gravity at large comoving wavenumber, 
the dispersion relation for acoustic modes is easily derived; the 
medium supports stable acoustic oscillations modified by radiative 
(``Silk'') damping on small scales (Silk 1967, 1968; Peebles \& 
Yu 1970; Weinberg 1971; Ma \& Bertschinger 1995; Hu \& Sugiyama 
1995ab, 1996, 1997; Dodelson 2003).  In the WKB approximation the 
damping term enters as a complex correction to the acoustic mode 
frequencies that is $\propto k^2/\dot{\tau}$, becoming increasingly 
important at small spatial scales (see, e.g., Dodelson 2003, \S8.4).

Most analytic calculations of radiative damping of acoustic modes 
neglect gravity at large comoving wavenumber, $k$, because the 
frequency of acoustic oscillations ($\omega\approx c_s k$, where 
$c_s$ is the sound speed) is large and the medium is Jeans stable.
Here, I show that on scales {\it smaller} than the Jeans
length the medium is linearly unstable to a slow diffusive mode
that is of order $\dot{\tau}^{-1}$.  Thus, on small scales where 
the medium is dynamically Jeans stable and supports stable (but 
radiatively damped) acoustic oscillations, it is unstable to an 
orthogonal diffusive mode.  This mode was earlier  
discussed by Yamamoto et al.~(1998).
The physics of this slow diffusive mode is already included in CMB 
codes like {\tt CMBFAST} and {\tt CAMB}, which include gravity on small 
scales in the tight-coupling limit in their solution to the coupled equations
for perturbations, and in earlier work employing full numerical 
solutions to the coupled perturbation equations.  
However, the physics of this mode
has been often neglected in analytic studies of the growth
of perturbations, and for this reason a more complete discussion 
is warranted; it may affect analytic estimates for the 
characteristic baryon-acoustic oscillation (BAO) scale, 
the physics of low-mass dark matter halos, and the transfer 
function.

In \S\ref{section:acoustic}, I review the physics of acoustic 
modes and radiative damping in the cosmological context, and I 
show that the unstable diffusive mode appears in a simple WKB 
treatment that includes gravity on small scales. The discussion 
is related to that of the ``terminal velocity'' mode 
in Yamamoto et al.~(1998).  I emphasize 
that the growth of this mode is not tied to the breakdown 
in the coupling between gas and radiation;  indeed, it appears
at the same order in $\dot{\tau}$ as Silk damping, and a temperature
difference between the radiation and the gas is not required for 
growth.  I note that analogous instabilities
should operate in the epoch of neutrino decoupling, or 
whenever the medium is partially supported against self-gravity
by a relativistic and diffusing particle species.  Section 
\ref{section:background} presents a complimentary analysis in 
an expanding background closest in spirit
to Silk (1967), and to the discussion of adiabatic modes in 
standard textbooks. In \S\ref{section:discussion}, I provide a brief summary. 

\section{Acoustic Modes \& Silk Damping}
\label{section:acoustic}

Following Hu \& Sugiyama (1996), the equations for the evolution of 
the moments of the photon temperature 
distribution $\Theta_n\,\,\,(n=0,1,2)$ in the 
tight-coupling limit are
\beqa
\dot{\Theta}_0+k\Theta_1+\dot{\Phi}\,&=&\,0 \\
\dot{\Theta}_1+k\left(\frac{2}{3}\Theta_2-\frac{1}{3}\Theta_0\right)-
\dot{\tau}\left(\Theta_1-\frac{iv_b}{3}\right)-\frac{k\Psi}{3}\,&=&\,0 \\
-\frac{2k}{5}\Theta_1-\frac{9}{10}\dot{\tau}\Theta_2\,&=&\,0,
\label{temp}
\eeqa
where $v_b$ is the baryon velocity and
the last expression neglects $\dot{\Theta}_2$.
The continuity and Euler equations for the baryons and dark matter are
\beqa
\dot{\delta}_b+ikv_b+3\dot{\Phi}\,&=&\,0  \\
\dot{v}_b+\frac{\dot{a}}{a}v_b-\frac{\dot{\tau}}{R}\left(v_b+
3i\Theta_1\right)+ik\Psi\,&=&\,0,  \\
\dot{\delta}_m+ikv_m+3\dot{\Phi}\,&=&\,0,  \\
\dot{v}_m+\frac{\dot{a}}{a}v_m+ik\Psi\,&=&\,0,
\eeqa
respectively,
where $R=3\rho_b/4\rho_r$, 
\beq
\dot{\tau}=d\tau/d\eta=-n_e\sigma_Ta\approx-\rho_b\kappa_Ta,
\label{taudot}
\eeq
$\eta$ denotes conformal time, $\kappa_T$ is the Thomson opacity,
and the notation of Dodelson (2003) has been adopted.
The field equations can be written as
\beq
k^2\Phi+3\frac{\dot{a}}{a}
\left(\dot{\Phi}-\Psi\frac{\dot{a}}{a}\right)
= 4\pi G a^2\left(\rho_b\delta_b+\rho_m\delta_m+4\rho_r\Theta_0\right) 
\eeq
and
\beq
k^2(\Phi+\Psi) =-32\pi Ga^2\rho_r\Theta_2,
\eeq
where $\rho_b$, $\rho_m$, and $\rho_r$ denote the baryon, matter, and radiation
energy density, respectively.  The approximate equality in equation (\ref{taudot})
assumes that one consider epochs sufficiently before recombination that the 
universe is fully ionized.
The above set of equations neglects moments of the temperature distribution
above the quadrupole, the polarization hierarchy, and neutrinos.
  
In the simplest derivation of Silk damping, one neglects both the 
gravitational terms and the expansion of the universe
(so that $\dot{a}/a={\dot{\Phi}}=\Phi=\Psi=0$) as well as the 
matter.   Expanding the time dependence as $e^{i\omega t}$, 
one finds that 
\beq
\omega\left[\omega^3+i\dot{\tau}\omega^2\left(1+\frac{1}{R}+\frac{8k^2}{27\dot{\tau}^2}\right)
-\omega\frac{k^2}{3}\left(1+\frac{8}{9R}\right)-i\dot{\tau}\frac{k^2}{3R}\right]=0.
\label{dispersion_nograv}
\eeq
Expanding to lowest order in the limit of large $\dot{\tau}$, there are
two acoustic modes
\beq
\omega_{\rm acoustic}
\approx \pm c_s k - \frac{ik^2}{2(1+R)\dot{\tau}}\left(c_s^2R^2+\frac{8}{27}\right)
\label{acoustic_mode}
\eeq
where $c_s^2=c^2/[3(1+R)]$, and the purely damped mode
(see, e.g., Blaes \& Socrates 2003)\footnote{Note that $\omega_{\rm damped}$
disappears from the analysis if the gas and radiation temperatures
are assumed to be equal, as in \S\ref{section:background} (e.g., 
Blaes \& Socrates 2003; Kaneko \& Morita 2006; Appendix B of Thompson 2008).}
\beq
\omega_{\rm damped}\approx-i\dot{\tau}\left(\frac{1+R}{R}\right)
-\frac{ik^2R}{27\dot{\tau}}\frac{(8-R)}{(1+R)^2}.
\label{diffusion_mode}
\eeq
Because $\dot{\tau}$ is always negative and
because I have written the time dependence as $e^{i\omega t}$, 
all terms in all three modes involving $\dot{\tau}$ represent 
damping.  Note that equation (\ref{dispersion_nograv}) contains 
a fourth mode that has $\omega=0$, which I now discuss when
gravity is included in the system.

\subsection{The Simplest Case with Gravity}

The neglect of gravity in the above derivation is justified
on small scales for acoustic modes whose frequencies are 
$\propto c_s k$ (eq.~[\ref{acoustic_mode}]).  It turns out 
that it is also justified for 
the purely damped radiative mode (eq.~[\ref{diffusion_mode}]).
However, including gravity on small scales introduces a
qualitatively different behavior to the fourth mode, which
is purely damped on scales larger than the Jeans length and 
purely unstable on scales smaller than the Jeans length.

Keeping the field equations, but again ignoring terms 
$\propto \dot{a}/a$, $\propto \dot{\Phi}$, and the equations 
for the matter, I find 
\beqa
\omega^4&+&i\dot{\tau}\omega^3\left[1+\frac{1}{R}+
\frac{8}{27\dot{\tau}^2}\left(k^2+\xi_r^2\right)\right] \nonumber \\
&-&\omega^2\left[\frac{k^2}{3}\left(1+\frac{8}{9R}\right)-
\frac{\xi_r^2}{27R}\left(9R^2+R-8\right)\right] \nonumber \\
&-&i\dot{\tau}\left(\frac{1+R}{R}\right)\omega\left[\frac{k^2}{3(1+R)} 
-\frac{\xi_r^2}{3}(1+R)-
\frac{8k^2R^2\xi_r^2}{81\dot{\tau}^2(1+R)}\right] \nonumber \\
&-&\frac{1}{9}k^2R\xi_r^2=0,
\label{dispersion_gravity}
\eeqa
where $\xi_r^2=16\pi G a^2 \rho_r$.
As before, 
there are four orthogonal modes admitted by equation (\ref{dispersion_gravity})
The first and second are gravity- and radiative-diffusion-modified 
acoustic modes.   To first order in $\dot{\tau}^{-1}$ these are
\beqa
\hspace*{-.5cm}
\omega_{\rm acoustic}&=&\pm\left[c_s^2k^2-\xi_r^2\frac{(1+R)}{3}\right]^{1/2} 
+\frac{i}{\dot{\tau}}\left(1-\frac{(1+R)\xi_r^2}{3c_s^2k^2}\right)^{-1}\nonumber\\
&\times&\left[\frac{-k^2}{2(1+R)}\left(c_s^2R^2+\frac{8}{27}\right)
+\frac{4\xi_r^2}{27}\left(R+\frac{(1+R)\xi_r^2}{3c_s^2k^2}\right)\right] 
\label{acoustic_gravity}
\eeqa
This expression should be compared with equation (\ref{acoustic_mode}).
Note that taking either the $k\rightarrow\infty$ or the $\xi_r^2\rightarrow0$
limits in equation (\ref{acoustic_gravity}) recovers equation (\ref{acoustic_mode}).
Furthermore, the first term in equation (\ref{acoustic_gravity}) is 
identical to what one would expect for the Jeans instability.  Setting the 
first term in square brackets to zero, I derive the Jeans length:
\beq
\frac{\lambda_J}{2\pi}=\left(\frac{k_J}{a}\right)^{-1}=\left(\frac{3c_s^2}{16\pi G\rho_r(1+R)}\right)^{1/2}.
\label{jeans}
\eeq
The third mode admitted by equation (\ref{dispersion_gravity}) 
is the strongly damped diffusion mode, $\omega_{\rm damped}$, of equation 
(\ref{diffusion_mode}), which is unmodified by the inclusion of gravity 
to second order in $\dot{\tau}^{-1}$.  The fourth
and final mode admitted by equation (\ref{dispersion_gravity}) is purely 
imaginary.  It is 
\beq
\omega_{\rm KH}\approx\frac{ik^2}{9\dot{\tau}}\left(\frac{R^2}{1+R}\right)
\left(\frac{\xi_r^2}{c_s^2k^2-\xi_r^2(1+R)/3}\right)
\label{simple}
\eeq
to first order in $\dot{\tau}^{-1}$.  This expression follows from equating 
the last two terms in equation (\ref{dispersion_gravity}). 
Note that in the limit that the medium is Jeans {\it stable} $c_s^2k^2> (1+R)\xi^2_r/3$ 
(large comoving wavenumber, small spatial scale)
equation (\ref{simple}) implies that the medium is unstable.
Conversely, on scales larger than $\lambda_{\rm J}$, $\omega_{\rm KH}$ is 
damped.  In the high$-k$ limit ($c_s^2k^2\gg (1+R)\xi^2_r/3$), 
equation (\ref{simple}) can be written simply as 
\beq
\frac{\omega_{\rm KH}}{a}\approx i\frac{R^2\xi_r^2}{3\dot{\tau}a}
=-i\frac{3\pi G}{\kappa_T c}\left(\frac{\rho_b}{\rho_r}\right),
\label{unstable_gas}
\eeq
which is independent of spatial scale.
Precisely when the medium is stable to the classical Jeans instability
it is unstable to a slow diffusive mode.

The characteristic frequency is smaller than that for any of the other 
modes: the timescale for growth is 
$[3\pi G/(\kappa_T c)]^{-1}\approx2\times10^{16}$\,s,
much longer than the radiation acoustic timescale,
$\omega_{\rm acoustic}^{-1}\sim(c_s k)^{-1}$, 
on a scale of order the Hubble radius and the 
interaction timescale $\omega_{\rm damped}^{-1}$. 
The subscript ``KH'' is used to emphasize the
connection with the Kelvin-Helmholtz timescale 
for gravitational contraction of a radiation pressure
supported self-gravitating body (see \S\ref{section:dm}).
The timescale $t_{\rm KH}$ is roughly $m_e/m_p$ shorter than the 
timescale for the radiative instability 
derived by Gamow (1949) and Field (1971) ($\sim10^{12}$\,yr;
their eqs.~20 \& 58, respectively),
and does not formally require a temperature difference
between the radiation field and the gas (\S\ref{section:background}).

\subsection{The Simplest Case with Dark Matter}
\label{section:dm}

The slow diffusive mode derived above changes when dark matter is 
included primarily because the strength of the potential changes.
In the WKB limit, the mode structure becomes more complicated
when the Euler and continuity equations for the dark matter
are included and the dispersion relation is cumbersome.
Nevertheless, neglecting $\dot{a}/a$, $\dot{\Phi}$ and neutrinos
as before, the slow diffusive mode can be isolated in the 
dispersion relation.  To first order in $\dot{\tau}^{-1}$, it reads
\beq
\omega_{\rm KH}\approx i\frac{k^2}{3\dot{\tau}}\left(\frac{R}{1+R}\right)
\left[\frac{\xi_b^2+\xi_m^2(1+8/(9R))}{c_s^2k^2-\xi_r^2(1+R)/3-\xi_m^2}\right]
\label{unstable_dm_full}
\eeq
where $\xi_{b,m,r}^2=16\pi G a^2\rho_{b,m,r}$. Compare equation (\ref{unstable_dm_full})
with equation (\ref{simple}).
Again taking the strongly Jeans {\it stable} limit of 
small spatial scales and large $k$ ($c_s^2k^2\gg\xi_m^2+\xi_r^2(1+R)/3$),
I have that
\beq
\omega_{\rm KH}
\approx i\frac{R\xi_m^2}{\dot{\tau}}\left[\frac{\rho_b}{\rho_m}+1+\frac{8}{9R}\right],
\label{unstable_r}
\eeq
In the limit of small $R$ this expression becomes 
\beq
\frac{\omega_{\rm KH}}{a}\approx -i\frac{128\pi G}{9\kappa_T c}
\left(\frac{\rho_m}{\rho_b}\right).
\label{unstable_dm}
\eeq
As in equation (\ref{unstable_gas}),
this timescale is independent of spatial scale.  
As long as the medium 
is diffusive (the tight-coupling approximation holds) and one 
consider scales smaller than the Jeans length, the medium is 
unstable. Furthermore, note that 
taking instead the large $R$ limit in equation (\ref{unstable_r}) 
recovers the basic form of equation (\ref{unstable_gas}), but 
with the substitution $\rho_b\rightarrow \rho_b+\rho_m$
in the numerator. However, for $z>1000$, $R<1$
and equation (\ref{unstable_dm}) is the relevant limit.
The characteristic growth timescale is then
\beq
t_{\rm KH}\,a\sim\frac{a}{\omega_{\rm KH}}\approx
\frac{9\kappa_T c}{128\pi G}\left(\frac{\rho_b}{\rho_m}\right)
\sim 4\times10^{14}\left(\frac{10\rho_b}{\rho_m}\right)
\,\,\,{\rm s}.
\label{unstable_scale}
\eeq
Note the correspondence with the Salpeter timescale for 
black hole growth.
The characteristic timescale in equation (\ref{unstable_scale}) 
can be understood as the timescale for 
a region of differential volume to radiate away its total
(relativistic) thermal energy content ($e\sim\rho_b  c^2$) at the 
Eddington limit 
($\dot{e}\sim4\pi G\rho_{\rm tot}c/\kappa_T$):\footnote{
The ``9'' in equations (\ref{unstable_dm}) \& 
(\ref{unstable_scale}) originates in the numerical
factor in equation (\ref{temp}) connecting $\Theta_1$ and $\Theta_2$
(see, e.g.,  Hu \& Sugiyama 1996).}
\beq
t_{\rm KH}\sim \frac{e}{\dot{e}}
\sim\frac{c\kappa_T}{4\pi G }\left(\frac{\rho_b}{ \rho_{\rm tot} }\right).
\label{unstable_edd}
\eeq
This is simply the Kelvin-Helmholz timescale.  $t_{\rm KH}$
can equivalently be thought of as the ratio of the square of 
the dynamical timescale, $(4\pi G\rho_{\rm tot})^{-1}$, to the 
photon-baryon interaction timescale, $(\kappa_T\rho_b c)^{-1}$. 
Alternatively, one may think of the growth timescale as 
$t_{\rm KH}=t_{\rm diff}\left(t_{\rm dyn}/t_{\rm r}\right)^2$,
where $t_{\rm diff}=ck^2/(3\kappa_T\rho_b)$ is the diffusion
time,  $t_{\rm dyn}=(4\pi G\rho_{\rm tot})^{-1/2}$ is the 
dynamical time, and $t_{\rm r}=(c_s k)^{-1}$ is the radiation
acoustic sound-crossing timescale on a scale $k^{-1}$.
Finally, one can also obtain this characteristic timescale
by equating the frictional force between the radiation and
the fluid due to Thomson scattering with the gravitational
force.  The timescale for instability is $\sim m_e/m_p$
shorter than the timescale for the instability discussed
in Gamow (1949) (see also Field 1971).  Note that 
because the mode that becomes unstable has $\omega=0$
in the adiabatic limit, one may identify this unstable mode
as the entropy mode (e.g., Lithwick \& Goldreich 2001).

Importantly, $t_{\rm KH}$ in equation (\ref{unstable_scale})
is roughly 100 times the age of the universe
at decoupling and approximately 500 times the age of the 
universe at radiation-matter equality.  At earlier times
the instability also operates, but because $t_{\rm KH}$
is a constant in the large-$k$ limit, it becomes increasingly 
long compared to the Hubble time at higher redshift.  This 
is most clearly seen by writing
\beq
t_{\rm KH}H(z)a\sim\frac{1}{\Omega_m}\left(\frac{\Gamma}{H(z)}\right),
\label{omegah}
\eeq
where $\Gamma$ is the interaction timescale and $H$ is the Hubble 
parameter.  Because $\Gamma\gg H(z)$ in the tight-coupling limit,
$t_{\rm KH}H(z)a$ is much larger than unity, and the instability is slow.

For a non-relativistic
discussion of this unstable mode and its potential astrophysical
applications, see Kaneko \& Morita (2005) and Thompson (2008).
The same mode has been discussed by Yamamoto et al.~(1998),
where they term it the ``terminal velocity mode'' (their Section 3.3),
since one can write the baryon velocity as $v_b\simeq k\Psi R/\dot{\tau}$
(their notation) by equating the gravitational force with the
drag force associated with the coupling to the radiation field.
As I have shown above, the mode is unstable at arbitrarily 
large (but, not infinite) $\dot{\tau}$, {\it with fixed growth timescale}, and operates 
until decoupling.  As Yamamoto et al.~(1998) show, the instability
smoothly joins the pressure-less collapse solution after decoupling.

Note that physically equivalent modes should operate 
whenever the medium is supported against self-gravity by a particle 
species that diffuses. For example, in the epoch before neutrino-matter
decoupling at a time $t_H\sim$1\,s and at a temperature larger than
$\sim10^{10}$\,K a similar mode is expected to 
have growth rate of order
\beq
\frac{\omega}{a}\sim -i\frac{4\pi G\rho_m}{\Gamma},
\eeq
where here $\rho_m$ is the {\it mass density} in all matter, and 
$\Gamma\sim G_F^2 T^5/\hbar$ is the weak interaction rate;
$\Gamma\approx0.2T_{\rm MeV}^5$ s$^{-1}$ where $T_{\rm MeV}=T/$MeV.  
The mass density 
in matter at $z\sim10^{10}$ is of order $\rho_m\sim 1$\,g cm$^{-3}$
and so I find that $(\omega/a)\sim -i\,10^{-6}$ s$^{-1}$.  That is,
the growth timescale is of order one million times the age of the
universe at that epoch.

\section{Slow Gravitational Instability \\ in an Expanding Background}
\label{section:background}

The preceding analysis approximated the background state as fixed.
However, the growth timescale derived exceeds the timescale for the background
state to change by a factor of more than 100.  This invalidates the 
WKB calculation and necessitates a treatment with an evolving
background.  This is most simply treated on small scales where
the gravitational potential may be approximated as Newtonian.
For simplicity of presentation, I treat the non-relativistic
case with $aT^4\ll \rho c^2$ and I neglect dark matter.  
In this case, the Eulerian equation governing the evolution 
in an inertial frame are  (e.g., Silk 1967)
\beq
\left(\frac{\p\rho}{\p t}\right)_{\bf r}+{\bf \bnabla_r\cdot}(\rho{\bf u})=0,
\label{continuity}
\eeq
\beq
\left(\frac{\p{\bf v}}{\p t}\right)_{\bf r}+{\bf u\cdot \bnabla u}
=-\frac{1}{\rho}{\bf \bnabla_r}p-{\bf \bnabla_r}\Phi,
\label{momentum}
\eeq
\beq
\left(\frac{\p e}{\p t}\right)_{\bf r}+{\bf u\cdot \bnabla_r}e+
\frac{4}{3}e{\bf \bnabla_r\cdot  u}=-{\bf \bnabla_r\cdot F},
\label{energy}
\eeq
\beq
\bnabla^2\Phi = 4\pi G\rho,
\label{gravity}
\eeq
and
\beq
{\bf F}=-\frac{c}{3\kappa_T\rho}{\bf \bnabla_r}e.
\label{diffusion}
\eeq
Here, $\rho$ is the mass density of gas, ${\bf u}$ is the velocity,
and $p=aT^4/3$ and $e=3p$ are the radiation
pressure and energy density, respectively.  The gas and radiation are assumed to 
have the same temperature, and it is assumed that $p$ greatly exceeds $p_g=n k_BT$,
the gas pressure.\footnote{For the general case with gas pressure, see
Thompson (2008).} 

In equations (\ref{continuity})-(\ref{diffusion}), I have
adopted the notation of Peebles (1993) (Chapter 5). 
The subscript ${\bf r}$ reflects reference to an inertial
coordinate system.  Transforming to a comoving coordinate
system with ${\bf x}={\bf r}/a(t)$ and 
${\bf u}=\dot{a}{\bf x}+{\bf v}({\bf x},t)=\dot{a}{\bf x}+a\dot{{\bf x}}$, 
where ${\bf v}$ is the peculiar velocity, and writing 
$\rho=\rho(t)(1+\delta({\bf x},t))$, with $\rho(t)\propto a(t)^{-3}$,
and taking $\ddot{a}/a=-(4/3)\pi G\rho(t)$,
equations (\ref{continuity})-(\ref{diffusion}) become
\beq
\frac{\p\delta}{\p t}+\frac{1}{a}{\bf \bnabla\cdot}[(1+\delta){\bf v}]=0,
\label{continuityc}
\eeq
\beq
\frac{\p{\bf v}}{\p t}+\frac{\dot{a}}{a}{\bf v}+
\frac{1}{a}{\bf v\cdot \bnabla v}
=-\frac{1}{a\rho(1+\delta)}{\bf \bnabla}p-\frac{1}{a}{\bf \bnabla}\phi,
\label{momentumc}
\eeq
\beq
\frac{\p e}{\p t}+\frac{1}{a}{\bf \bnabla\cdot}(e{\bf v})+
4\frac{\dot{a}}{a}e+\frac{p}{a}{\bf \bnabla\cdot  v}
=\frac{c}{3\kappa\rho a^2}{\bf \bnabla\cdot}\left[\frac{1}{(1+\delta)}{\bf \bnabla}e\right],
\label{energyc}
\eeq
\beq
\bnabla^2\phi = 4\pi G a^2 \delta,
\label{gravityc}
\eeq
where time and space derivatives are now understood to be
in the comoving frame and $\Phi=\phi({\bf x},t)+(2/3)\pi G\rho(t)a^2x^2$. 
In analogy with the density perturbation I write  $e=e(t)(1+\varepsilon({\bf x},t))$,
and equation (\ref{energyc}) can be rewritten as
\beq
\frac{\p \varepsilon}{\p t}+
\frac{1}{a}{\bf \bnabla\cdot}\left[(1+\varepsilon){\bf v}\right]+
\frac{1+\varepsilon}{3a}{\bf \bnabla\cdot  v}
=\frac{c}{3\kappa\rho a^2}{\bf \bnabla\cdot}
\left[\frac{{\bf \bnabla}\varepsilon}{1+\delta}\right].
\label{energyc2}
\eeq
So far the equations written are completely general for the 
problem at hand.  I now specialize to the case of small departures
from the background state so that terms of order $\delta |v|$, $\delta\varepsilon$,
and $\varepsilon|v|$ can be neglected.  
The resulting equations can be written as
\beq
\frac{\p^2\delta}{\p t^2}+2\frac{\dot{a}}{a}\frac{\p\delta}{\p t}
-4\pi G\rho \delta=\frac{e}{3\rho a^2}\bnabla^2\varepsilon
\label{wave}
\eeq
and
\beq
\frac{\p \varepsilon}{\p t}-
\frac{4}{3}\frac{\p\delta}{\p t}
=\frac{c}{3\kappa\rho a^2}\nabla^2\varepsilon.
\label{diffusions}
\eeq
Note that the left hand side of equation (\ref{diffusions})
is simply
\beq
\frac{\p \varepsilon}{\p t}-
\frac{4}{3}\frac{\p\delta}{\p t}=
\frac{4}{3}\frac{\p}{\p t}\left(\frac{\Delta s}{s}\right),
\eeq
where $s$ is the total entropy.
Writing $\varepsilon=\Delta e/e=4\Delta T/T$, equations
(\ref{wave}) and (\ref{diffusions}) become identical to 
equations (1) \& (2) of Silk (1967).
Expanding these expressions in Fourier components with
wavenumber $\tilde{k}=k/a$, taking $a(t)\propto t^{2/3}$,
and defining the temperature perturbation as  $q=\varepsilon/4$,
results in two coupled ordinary differential equations for the 
time evolution of the density and temperature perturbations:
\beq
\ddot{\delta}_{\tilde{k}}+\frac{4}{3t}\dot{\delta}_{\tilde{k}}
-\frac{2}{3t^2}\delta_{\tilde{k}}+3c_r^2\tilde{k}^2\,q_{\tilde{k}}=0
\label{delta_evol}
\eeq
and
\beq
\dot{q}_{\tilde{k}}-
\frac{1}{3}\dot{\delta}_{\tilde{k}}
+\tilde{\omega}\,q_{\tilde{k}}=0,
\label{diffusions_1}
\eeq
where over-dots denote time derivatives, 
\beq
\tilde{\omega}=\frac{c\tilde{k}^2}{3\kappa\rho}
\eeq
is the diffusion rate,
\beq
c_r^2=\frac{4}{9}\frac{e}{\rho}=\left.\frac{\partial p}{\partial \rho}\right|_s
\eeq
is the adiabatic radiation pressure sound speed,
and the subscript $\tilde{k}$ has been added to 
emphasize that these equations are for the 
evolution of the density and temperature, $\delta$ and $q$, 
at a  specific comoving wavenumber.

\subsection{Adiabatic Modes}

In the adiabatic limit $\tilde{\omega}=0$ in equation
(\ref{diffusions_1}), 
$\partial/\partial t(\Delta s/s)=0$, $q=(1/3)\delta$,
and the system becomes (e.g., Peebles eq.~5.124)
\beq
\ddot{\delta}_{\tilde{k}}
+\frac{4}{3t}\dot{\delta}_{\tilde{k}}
+\left(c_r^2\tilde{k}^2-\frac{2}{3t^2}\right)\delta_{\tilde{k}}=0.
\label{peebles}
\eeq
In the long-wavelength limit ($\tilde{k}\rightarrow0$)
one recovers the classic equation for the time evolution of 
the density contrast, modeled as a purely pressure-less fluid,
which has the growing and damped solutions 
$\delta_+\propto t^{2/3}$ and $\delta_-\propto t^{-1}$, respectively.
Taking $\tilde{k}\rightarrow\infty$, equation (\ref{peebles})
describes an acoustic wave, damped by the expansion of the 
universe.

\begin{figure*}
\centerline{\psfig{file=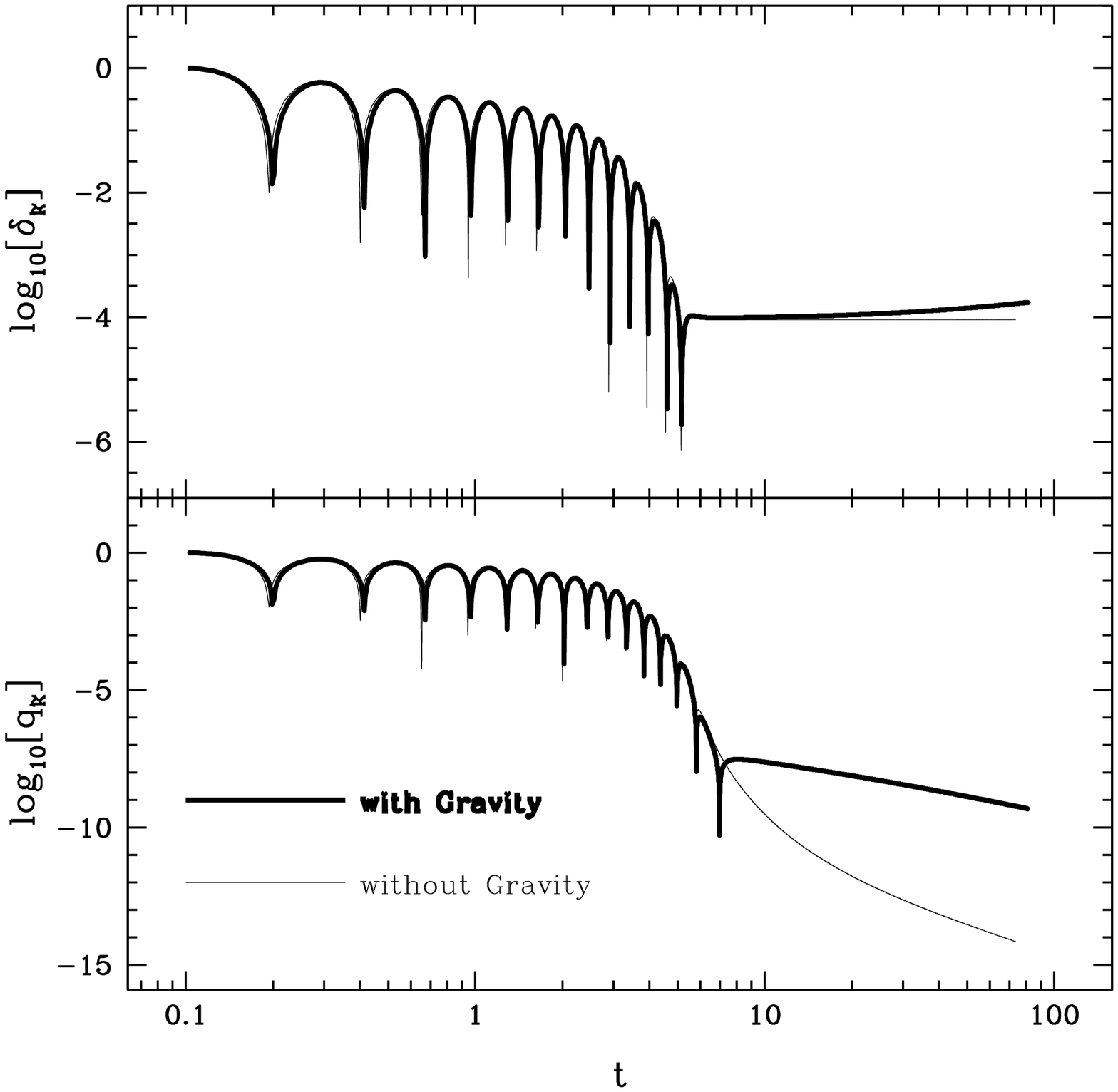,width=8.6cm}\psfig{file=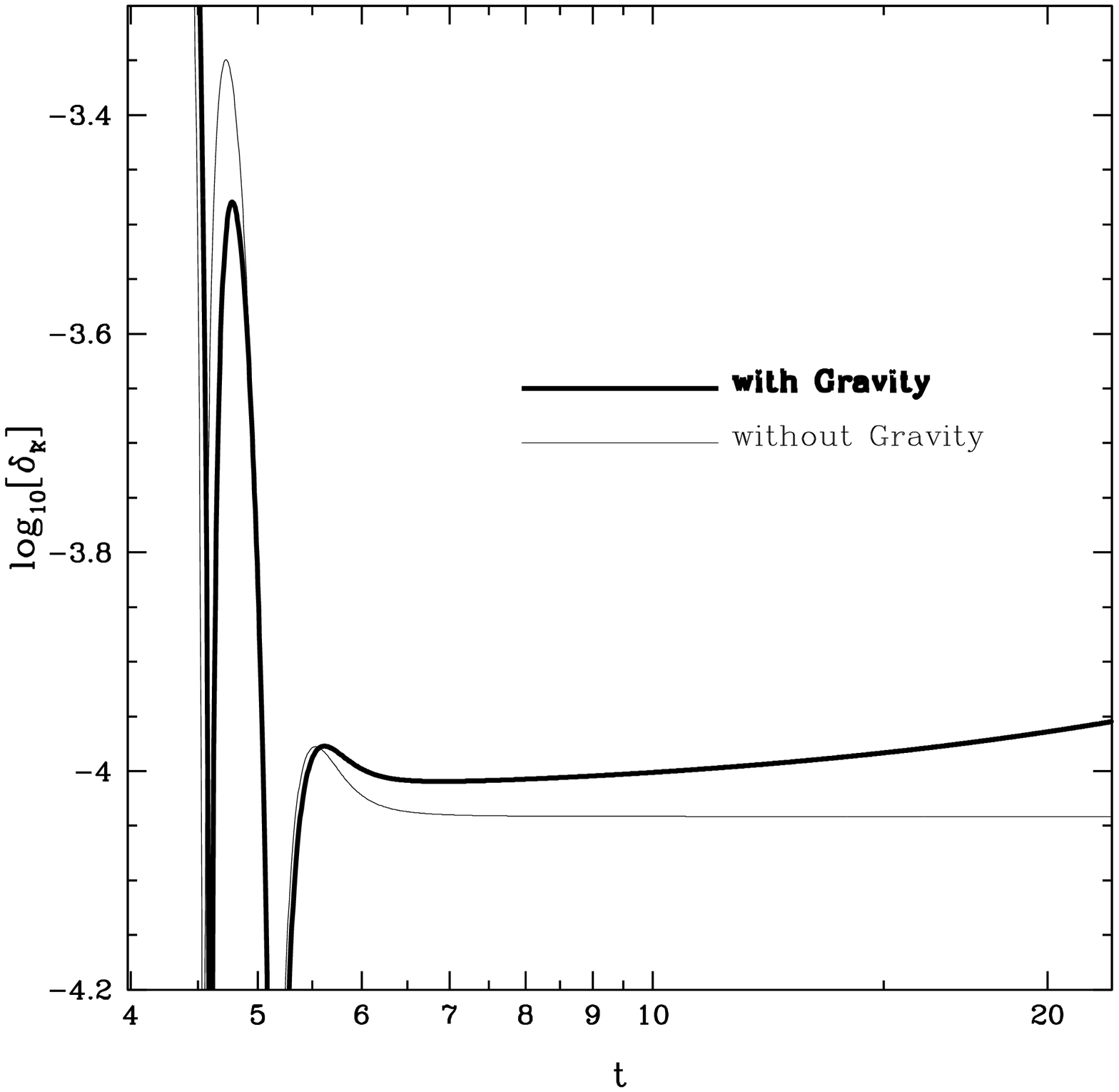,width=8.6cm}}
\figcaption[damp]{{\it Left Panel:} Solution to equations (\ref{delta_evol}) and (\ref{diffusions_1})
at large $\tilde{k}$ as a function of time both with gravity (the ``$2/3t^2$'' term; heavy solid line) 
and without gravity (light solid). The top and bottom panels show the 
evolution of the density and temperature perturbations, $\delta_{\tilde{k}}(t)$ and $q_{\tilde{k}}(t)$.  
The acoustic wave is rapidly damped by both the expansion of the universe and 
diffusive damping.   At early times the two solutions track each other closely.
After the mode is damped, the evolution is qualitatively different as a result
of the slow diffusive mode. {\it Right Panel:} A zoom in on the late-time evolution 
of $\delta_{\tilde{k}}$ to highlight the difference between the calculations with and 
without gravity.  The increase in $\delta_{\tilde{k}}(t)$ at late times 
is not a result of the breakdown of the tight-coupling approximation, but is
instead a result of the instability highlighted in this paper.
Importantly, in a more complete calculation the medium 
becomes optically-thin, the radiation decouples, and the solution joins the
classic pressure-less free-fall collapse solution.  See Yamamoto et al.~(1998)
and Singh \& Ma (2002) for detailed calculations at small scales.
\vspace{.7cm}
\label{fig:damp}}
\end{figure*}

\subsection{Non-Adiabatic Evolution}

The full non-adiabatic evolution is more complicated.  First, as shown by
Silk (1967) and others, the acoustic modes are damped
by radiative diffusion.  Second, --- the purpose of this 
paper --- as shown in the 
previous sections, it is formally incorrect to neglect
gravity at large comoving wavenumber, because to do so is
to effectively set the driving term for the diffusive
 gravitational instability to zero. 

To illustrate the importance  of these effects, I solve
equations (\ref{delta_evol}) and (\ref{diffusions_1})
for $\delta_{\tilde{k}}(t)$ and $q_{\tilde{k}}(t)$
with $\dot{\delta}_{\tilde{k}}(t=0)=0$, $\delta_{\tilde{k}}(t=0)=1$, 
$q_{\tilde{k}}(t=0)=1/3$, $c_r^2\tilde{k}^2=100/t^{2/3}$, and 
$\tilde{\omega}=t^2/\sqrt{10}$ over the range $0.1\leq t\leq100$.
These parameters ensure that the mode considered is much smaller
than the Jeans scale and that the diffusion rate is small.
Figure \ref{fig:damp} shows results for the time evolution of the density and 
temperature perturbations with and without gravity
(heavy and light solid lines, respectively).  The acoustic
mode is damped by both the expansion of the universe and 
Silk damping.  At early times, the evolution with and 
without gravity is very similar.  At later times, the 
presence of the unstable growing mode becomes more
pronounced and the evolution of $\delta_{\tilde{k}}$ and $q_{\tilde{k}}$ 
is qualitatively different.  The right panel shows a 
zoomed-in version of the late-time evolution of $\delta_{\tilde{k}}$.
At $t\approx6$, as the mode becomes fully damped, 
the fractional difference in $\delta_{\tilde{k}}$ between
the two solutions is a few to 10\%.  Of course, this example does 
not include the physics of decoupling of the mode and gravitational
free-fall into the dark matter potential, which would be included
in a full calculation, and so the difference between the 
two solutions becomes exaggerated at late times 
(see Yamamoto et al.~1998; Singh \& Ma 2002).

\section{Discussion}
\label{section:discussion}

The original treatments of acoustic modes near the epoch of
radiation-matter equality focused on the importance of radiative
damping of acoustic modes at large comoving wave number.  The 
general equations derived by Silk (1967) (and Silk 1968; Peebles \& Yu 1970; Weinberg 1971)
contain the physics described above, but the slow diffusive mode 
these equations admit was not discussed because gravity
was neglected when calculating the damping of acoustic modes
on small scales.  This approximation eliminated the unstable 
diffusive mode. Similarly, the work of
Hu \& Sugiyama (1996) provides a concise analytic treatment
of acoustic mode damping, but again neglects the importance 
of gravity at large comoving wavenumber.  
For the unstable mode I have highlighted, the $k$-dependence
drops out in the high-$k$ limit because of the $k$-dependence
of the diffusion timescale.  Thus, to first order in the limit
of large $\dot{\tau}$ it is not consistent to neglect gravity
at high-$k$. Yamamoto et al.~(1998) have discussed 
the same instability, but it is important to emphasize
that the timescale for the mode's growth is a constant on small
scales as long as the tight coupling approximation is valid 
(eq.~\ref{unstable_scale}). 
In this way, the mode's existence is not a manifestation of the breakdown
of the tight coupling approximation; even at very high $\dot{\tau}$, 
the instability exists with fixed $t_{\rm KH}a$.

Most importantly,  
on all scales smaller than the Jeans length and at all times
before decoupling, the medium is
unstable, albeit on a long timescale.   Thus, it is not formally 
correct to think of the medium on scales below the horizon as 
stable.  Although it is dynamically stable, it is unstable
on the Kelvin-Helmholtz timescale for a radiation-pressure supported
self-gravitating medium ($\sim\kappa_T c/G$).  The instability should operate at 
all scales smaller than the Jeans length for which the 
medium is tightly coupled.   The physics of the mode is simply
that self-gravitating radiation pressure supported media radiate 
their total internal energy content at the Eddington limit.  
It is qualitatively different than the mode described by 
Gamow (1949) and Field (1971).

The physics of this mode is already contained in the
popular codes {\tt CMBFAST} and {\tt CAMB} for calculating
the growth of perturbations in the tight coupling limit.
However, precision analytic and numerical
studies of primordial perturbations in density and temperature
that neglect gravity on small scales may find systematic small
differences with respect to more complete calculations that are 
attributable to the growth of this mode 
(e.g., in the transfer function [e.g., Yamamoto et al.~1998], 
or the BAO scale, or in the initial conditions for the 
formation of very small scale dark matter halos).

As noted in \S\ref{section:acoustic}, the 
instability identified in equation (\ref{unstable_scale})
should operate at essentially all times before decoupling
because it depends only on the ratio $(\rho_m/\rho_b)$ and 
the opacity $\kappa_T$.  However, because the growth
timescale is constant, it becomes increasingly long compared to 
the age of the universe at earlier times (see eq.~[\ref{omegah}]).  
Thus, one expects it to be most important at the latest times
for which the assumptions apply: the largest Jeans-stable modes 
at a time near decoupling.  Even here, the growth timescale is 
much longer than the age of the universe, as can be seen from 
equation (\ref{omegah}).  More generally, the medium should be 
unstable to analogous modes any time it is supported
against self-gravity by the radiation pressure provided by 
a diffusing particle species (\S\ref{section:dm}).

\acknowledgements

I thank Chris Hirata for a number of useful conversations, as well 
as David Spergel, Matias Zaldarriaga, Lam Hui, Scott Dodelson,
and Martin White for discussions.  I am 
additionally grateful to the Aspen Center for Physics, where a 
portion of this work was completed, and to D.~Zifkin for encouragement.
This work is supported in part by an Alfred P.~Sloan Fellowship.

\end{document}